\documentclass{JHEP3} 

\usepackage{epsfig,multicol}

\title{Lorentz and CPT violations from Chern-Simons modifications of QED}

\author{Alexander A. Andrianov\thanks{Permanent address: V.A.Fock Department of Theoretical Physics,
Sankt-Petersburg State University, 
198504 Sankt-Petersburg, Russia }, Paola Giacconi and Roberto Soldati\\
Dipartimento di Fisica, Universit\'a
di Bologna ,\\
Istituto Nazionale di Fisica Nucleare,
Sezione di Bologna,\\via Irnerio 46, 40126 Bologna, Italia\\
E-mail: \email{andrianov@bo.infn.it}, \email{giacconi@bo.infn.it}, \email{soldati@bo.infn.it}}

\abstract{ The possibility of a small
modification of spinor Quantum Electro-Dynamics is 
reconsidered, in which Lorentz and CPT non-covariant 
kinetic terms for photons and fermions are present.
The corresponding free field theory is carefully discussed.
The finite one-loop parity-odd induced effective action 
is unambiguously calculated using the physical cutoff method,
which manifestly encodes the maximal residual symmetry group
allowed by the presence of the Lorentz and CPT breaking axial-vector. 
This very same induced effective action, which
is different from those ones so far quoted in the Literature,
is also re-derived by means of the dimensional regularization, 
provided the maximal residual symmetry 
is maintained in the enlarged $D$-dimensional space-time.
As a consequence, it turns out that the requirement of
keeping the maximal residual symmetry at the quantum level
just corresponds to the physical renormalization prescription
which naturally fixes the one-loop parity-odd induced effective action.}


\preprint{DFUB/12/01}
\begin{document}
\section{Introduction}
So far the Lorentz symmetry has been proven to hold
with a very high accuracy. Nevertheless, one can inquire about
whether the Special Relativity Theory is, for some unknown reasons,
only approximate. The modern quantum field theoretical viewpoint
admits that the {\sl spontaneous} Lorentz symmetry breaking is not
excluded and, at this expense, the CPT symmetry can be also broken in a
local field theory.

The occurrence of a very small deviation from the Lorentz
covariance has been discussed recently \cite{1,2,3} within the
context of the Standard Model of electroweak interactions. There,
some background or cosmological fields are implied, leading
to deviations from Lorentz-covariant dispersion laws for the free propagation of certain
particles. The possibility
of a tiny breaking of the equivalence between different Lorentz
frames   has been reconsidered  and severe
astrophysical and laboratory bounds on it have been derived \cite{4,5}.

As the photon is a test particle of the Special Relativity the
most crucial probes concern  Lorentz and
CPT symmetry breaking modification of Quantum Electro-Dynamics.
A basic requirement is that such a modification of spinor QED does not spoil its fundamental 
character provided by renormalizability, unitarity and gauge invariance in the ordinary {\it 3+1}
dimensional Minkowski space-time. Within this framework, a certain realization of the
Lorentz and CPT symmetry breaking might be obtained
in terms of two kinds of additional CPT-odd kinetic terms in the action.
The first one, concerning photons, is a  Chern-Simons (CS) action
\cite{1} involving a constant four-vector $\eta_\mu$, whilst the second one is
a CPT-odd kinetic term for fermions, which might describe a coupling
to a constant axial-vector torsion-like background field $b_\mu$ \cite{2}.
Such an  extension of spinor QED
does not break the gauge symmetry of the action although it modifies the
dispersion relations for different polarizations of photons and Dirac's spinors \cite{1,2,6}.
In Section 2 this non-covariant modification of photon kinematics is  presented
in the axial gauge, which turns out to be the most natural 
gauge choice in the presence of a given constant four-vector.
In Section 3 we shall analyze the details of the dispersion law of free spinors and 
derive henceforth the fermion stability bound.

Ultimately, the present analysis of the Lorentz and CPT symmetry breaking is
parameterized by two constant four-vectors $\eta_\mu$ and $b_\mu$ which are not
necessarily collinear. Their actual dynamical origin  represents an
interesting problem to be tackled.

One of the possible ways to induce Lorentz and CPT symmetry breaking by a dynamical mechanism 
has been suggested recently \cite{7}. Namely,
the spontaneous breaking of the
Lorentz symmetry {\it via} the Coleman-Weinberg mechanism \cite{8} has
been proven for a class of models with the Wess-Zumino interaction
between abelian gauge fields and a pseudo-scalar massless Axion field $\theta(x)$
(AWZ models). In those models, the proportionality between the vacuum expectation value of
the gradient of the Axion field
$\left<\partial_\mu \theta\right>_0$ and the constant four-vector $ \eta_\mu $ might
be a natural implementation for Lorentz and CPT symmetry breaking \cite{9},
just relating its origin to the assumed existence of  {\sl quintessence} fields \cite{10}. 
The background vector $b_\mu$ might be also related 
to some constant background torsion $\epsilon_{\mu\nu\rho\sigma} T^{\nu\rho\sigma}$ 
in the large scale Universe \cite{11}. As well, an example of the fermionic superfluid 
system in which the CPT-odd CS terms are induced by chiral fermions
can be found in Ref.~\cite{12}. 

The presence of the background vector $b_\mu$ also leads 
to the radiatively induced
Chern-Simons term, {\it i.e.} it modifies the classical tree level magnitude 
of the four-vector $\eta_\mu$ for photons. However, the amount of
this effect has been disputed 
\cite{3},\cite{13}-\cite{20} 
and the firm prediction has not yet been
found at the formal level of the renormalization theory.

On the other hand, it turns out that in the modified QED under consideration
its dynamics, nonperturbative in axial vector $b_\mu$,  
allows for the decay of highly energetic fermions - with momenta above the
stability bound - into two fermions and an anti-fermion with 
lower energies \cite{2,6}. 
As we shall discuss in Section 4, the existence of this process suggests
that a global symmetry, which has to be realized at the quantum level in 
particular inertial frames, is that one generated by the 
{\sl maximal residual symmetry subgroup} which is the small group
of a specific background vector $b^\mu$. 
The requirement that the latter symmetry is realized at the quantum level
order-by-order in perturbation theory just leads  to the unique 
non-vanishing value of the radiatively induced CS coefficient, as the 
corresponding diagram becomes finite in a way dictated by the manifest 
invariance under the maximal residual symmetry subgroup. 

In Section 5 we  show that the very same non-vanishing coefficient of the
radiatively induced CS term can be derived
within the properly defined dimensional regularization scheme 
($\overline{DR}$) which supports the maximal residual space-time symmetry 
and thereby the same physical content, {\it i.e.} the same physical renormalization
prescription.

As a further important issue, 
one should analyse the consistency of the proposed CS modification of Quantum
Electro-Dynamics, in  attempt to describe
particle physics in a wide range of energies. 
In particular, the stability \cite{6} and microcausality \cite{19,21} 
issues have to be fully examined. In the Conclusion
we comment about their status and consequences and we argue that there 
is a window for the construction of a consistent 
Maxwell-Chern-Simons spinor QED.                                   
\section{Maxwell-Chern-Simons free fields in the axial gauge}
The Maxwell-Chern-Simons Lagrange density \cite{1} in the axial gauge for the
radiation field reads
\begin{equation}
{\cal L}_{\mbox{\rm MCS}}=-{1\over 4}F^{\nu\lambda}F_{\nu\lambda}-{1\over 2}\eta_\alpha
A_\beta\tilde F^{\alpha\beta}-B\eta_\alpha A^\alpha\ ,
\label{2.1}
\end{equation}
where $B(x)$ is the auxiliary field and $\tilde F^{\alpha\beta}\equiv
(1/2)\epsilon^{\alpha\beta\rho\sigma}F_{\rho\sigma}$.
The Euler-Lagrange field equations are
\begin{eqnarray}
\partial_\mu F^{\mu\nu}&=&\eta_\mu\tilde F^{\mu\nu}+B\eta^\nu \ ,\label{2.2a}\\
\eta^\alpha A_\alpha &=&0\ ,\label{2.2b}
\end{eqnarray}
or, equivalently, in terms of the gauge potentials
\begin{eqnarray}
&&\partial^2 A^\nu -\eta_\mu\epsilon^{\mu\nu\rho\sigma}\partial_\rho A_\sigma= 
\partial^\nu(\partial\cdot A)+\eta^\nu B\ ,\label{2.3a}\\
&&\eta^\alpha A_\alpha=0\ .\label{2.3b}
\end{eqnarray}
After contraction of eq.~(\ref{2.3a}) with $\partial_\nu$ we find
\begin{equation}
(\eta\cdot\partial)B(x)=0,\label{2.4}
\end{equation}
whence, if we require suitable boundary conditions at 
infinity\footnote{The inversibility of the differential operator
$(\eta\cdot\partial)$ is actually a legitimate requirement only for $\eta^\nu$
purely space-like \cite{22}.} for the
auxiliary field, we can set $B(x)\equiv 0$.

After contraction of eq.~(\ref{2.3a}) with $\eta^\nu$ we get
\begin{equation}
(\eta\cdot\partial)(\partial\cdot A)=0\ ,
\label{2.5}
\end{equation}
and, again, by imposing suitable boundary conditions at
infinity  we obtain $\partial\cdot A=0$. This means that, within the
axial gauge, the gauge potential of the radiation field contains only two
(eventually physical) field degrees of freedom.

Going in the momentum representation, {\it i.e.}
\begin{equation}
A_\nu (x)\equiv\int{d^4 k\over (2\pi)^{3/2}}\tilde A^\nu (k)\exp\{ik\cdot x\}\,
\label{2.6}
\end{equation}
the equations of motion take the form
\begin{eqnarray}
&&\left(k^2g^{\nu\sigma}-i\epsilon^{\mu\rho\nu\sigma}\eta_\mu
k_\rho\right)\tilde A_\sigma=0\ ,\label{2.7a}\\
&&\eta^\sigma\tilde A_\sigma(k)=0= k^\sigma\tilde A_\sigma(k)\ ,\label{2.7b}
\end{eqnarray}
In order to pick out the two independent field degrees of freedom, let us
introduce the useful quantity
\begin{equation}
\mbox{\tt D}\equiv (\eta\cdot k)^2- \eta^2k^2 \ ,
\label{2.8}
\end{equation}
and consider the projector onto the two-dimensional hyperplane orthogonal to
$\eta^\nu$ and $k^\nu$: namely,
\begin{equation}
\mbox{\rm e}^{\mu\nu}\equiv g^{\mu\nu}-\frac{\eta\cdot k}{\mbox{\tt D}}\left(\eta^\mu k^\nu
+\eta^\nu k^\mu\right)+\frac{k^2}{\mbox{\tt D}}\eta^\mu \eta^\nu + \frac{\eta^2}{\mbox{\tt D}}k^\mu k^\nu\ ,
\label{2.9}
\end{equation}
which fulfills
\begin{equation}
\mbox{\rm e}^{\mu\nu}\eta_\nu=\mbox{\rm e}^{\mu\nu}k_\nu=0\ ,\quad \mbox{\rm e}^\mu_{\
\lambda} \mbox{\rm e}^{\lambda\nu}=\mbox{\rm e}^{\mu\nu}\ ,\quad \mbox{\rm e}^\mu_{\ \mu}=2\ .
\label{2.10}
\end{equation}
We can always select two real orthonormal four-vectors corresponding to the
linear polarizations
in such a way that 
\begin{equation}
\mbox{\rm e}_{\mu\nu}=\ -\sum_{a=1,2}\mbox{\rm e}^{(a)}_\mu\mbox{\rm e}^{(a)}_\nu\ ,\quad
g^{\mu\nu}\mbox{\rm e}^{(a)}_\mu\mbox{\rm e}^{(b)}_\nu=\ -\delta^{ab}\ .
\label{2.11}
\end{equation}
It is also convenient to introduce another couple of four-vectors, which
describe left- and right-handed polarizations which in our case generalize the circular 
polarizations of the conventional QED. To this aim, let us first define
\begin{equation}
\epsilon^{\mu\nu}\equiv \mbox{\tt D}^{-1/2}\epsilon^{\mu\nu\alpha\beta}\eta_\alpha
k_\beta\ ,
\label{2.12}
\end{equation}
which enjoys
\begin{equation}
\mbox{\rm e}^{\mu\nu}=\epsilon^{\mu\lambda}\epsilon^\nu_{\ \lambda}\ ,\quad
\epsilon^{\mu\nu}=\mbox{\rm e}^{\mu\lambda}\epsilon_\lambda^{\ \nu}\ .
\label{2.13}
\end{equation}
Notice that we can always choose $\mbox{\rm e}_\mu^{(a)}$ in such a way to satisfy
\begin{equation}
\epsilon^{\mu\nu}\mbox{\rm e}_\nu^{(1)}=\mbox{\rm e}^{(2)\mu}\ ,\quad 
\epsilon^{\mu\nu}\mbox{\rm e}_\nu^{(2)}=\ -\mbox{\rm e}^{(1)\mu} \ .
\label{2.14}
\end{equation}
Let us now construct the two orthogonal projectors
\begin{equation}
P_{\mu\nu}^{(\pm)}\equiv{1\over 2}\left(\mbox{\rm e}_{\mu\nu}\pm
i\epsilon_{\mu\nu}\right)\ .
\label{2.15}
\end{equation}
and set, {\it e.g.},
\begin{eqnarray}
&&\varepsilon_\mu^{(L)}\equiv {1\over 2}\left(\mbox{\rm e}_\mu^{(1)}+i\mbox{\rm
e}_\mu^{(2)}\right)=P_{\mu\nu}^{(+)}\mbox{\rm e}^{(1)\nu}\ ,\label{2.16a}\\ 
&&\varepsilon_\mu^{(R)}\equiv {1\over 2}\left(\mbox{\rm
e}_\mu^{(1)}-i\mbox{\rm e}_\mu^{(2)}\right)=P_{\mu\nu}^{(-)}\mbox{\rm e}^{(1)\nu}
\ ,\label{2.16b}
\end{eqnarray}
in such a way that
\begin{equation}
\mbox{\rm e}_{\mu\nu}=\ -\sum_{a=L,R}\left\{
\varepsilon^{(a)}_\mu\varepsilon^{(a)*}_\nu +
\varepsilon^{(a)}_\nu\varepsilon^{(a)*}_\mu \right\}\ . 
\label{2.17}
\end{equation}
As a consequence, from eq.~(\ref{2.14}), we can readily check that
\begin{equation}
i\epsilon^{\mu\nu}\varepsilon_\nu^{(L)}=\varepsilon^{(L)\mu}\ ,\quad
i\epsilon^{\mu\nu}\varepsilon_\nu^{(R)}=\ -\varepsilon^{(R)\mu}\ .
\label{2.18}
\end{equation}
At this point we have to stress that the left- and right-handed (or chiral) polarizations only
approximately correspond to the circular ones of Maxwell QED \cite{23}. In the presence of CS kinetic
term the field strengths of electromagnetic wave are typically not orthogonal
to the wave vector \cite{1,7}. But just the above handedness is conserved and not the conventional
circular polarizations, so that this L,R chirality makes more physical sense in the
analysis of observational phenomena.

After this preliminary work on the polarizations
it is immediate to solve the field equations, at least in the 
case of $\eta_\mu$ purely space-like, {\it i.e.} $\eta_\mu=(0,\vec\eta)$. 
As a matter of fact, it turns out that in this case there 
are always four real solutions of the basic equation
\begin{equation}
(k^2)^2-\mbox{\tt D}=(k^2)^2-(\eta\cdot k)^2+\eta^2k^2=0\ ,
\label{2.19}
\end{equation}
which read
\begin{equation}
k^2_{0\pm}=\omega^2_\pm({\vec k},\vec\eta)=
{\vec k}^2+{1\over 2}\vec\eta^2\pm
|\vec\eta|\sqrt{{\vec k}^2\cos^2\varphi+{1\over 4}\vec\eta^2}\ ,
\label{2.20}
\end{equation}
where we have set ${\vec k}\cdot\vec\eta=|{\vec k}||\vec\eta|\cos\varphi$.
Then, from eq.~(\ref{2.7b}), we can
write 
\begin{equation}
\tilde A_\mu (k)=\sum_{a=L,R}\varepsilon^{(a)}_\mu (k) F_a (k)\ ,
\label{2.21}
\end{equation}
whence the solutions of eq.~(\ref{2.7a}) are easily found to be
\begin{equation}
F_L (k) = f_L (k)\delta[k_0^2 - \omega_+^2({\vec k},\vec\eta)]\ ,\quad
F_R (k) = f_R (k)\delta[k_0^2 - \omega_-^2({\vec k},\vec\eta)]\ ,
\label{2.22}
\end{equation}
where $f_{L,R}(k)$ are arbitrary functions though regular on the supports of
the corresponding $\delta$-distributions. 

According to the analysis of Refs.~\cite{1,2,9,21}, 
we see that monochromatic plane wave
solutions are possible only with a definite chiral polarization. For
instance,
in the space-like case $\eta^2 < 0$, after choosing the privileged frame in which
 $\eta_\mu=(0,\eta_1,0,0)$, we
find
\begin{equation}
\omega_\pm({\vec k},\eta_1)=\sqrt{|{\vec k}|^2+{1\over 2}\eta_1^2\pm
|\eta_1|\sqrt{k_1^2+{1\over 4}\eta_1^2}}\ ,
\label{2.23}
\end{equation}
which yields
\begin{equation}
\left|\vec\upsilon_{g\pm}\right|={1\over \omega_\pm}\sqrt{k_\perp^2 +k_1^2\left(
1\pm {|\eta_1|\over \sqrt{4k_1^2+\eta_1^2}}\right)^2}\ ,\quad
k_\perp^2=k_2^2+k_3^2\ , 
\label{2.24}
\end{equation}
and turns out to be always smaller than one.
This might lead to some observable
birefringence phenomenon, because the group velocities of the wave-packets
made out of chirally polarized CS-photons are always smaller than one.

On the contrary, in the time-like case $\eta^2 > 0$, after choosing the privileged frame in which
$\eta^\mu=(\eta_0,0,0,0)$, we obtain that left-handed and right-handed CS-photons
of a wave vector ${\vec k}$ travel with frequencies 
\begin{equation}
\omega_\pm({\vec k},\eta_0)=\sqrt{|{\vec k}|(|{\vec k}|\pm |\eta_0|)}\ ,
\label{2.25}
\end{equation} 
respectively. In this case the group velocities of the wave-packets
made out of chirally polarized CS-photons are 
\begin{equation}
\vec\upsilon_{g\pm}=\nabla_{{\vec k}}\omega_\pm({\vec k},\eta_0)=
{{\vec k}\over \omega_\pm}\left(
1\pm{|\eta_0|\over 2|{\vec k}|}\right)\ ,
\label{2.26}
\end{equation}
whence it immediately follows that $\left|\vec\upsilon_{g\pm}\right|$ is always bigger than one.
Moreover, for $|{\vec k}| < |\eta_0|$ one reveals imaginary energy solutions
(runaway solutions) which spoil the stability of such an electrodynamics.

The earlier analysis of the radiowaves patterns from distant galaxies had been
performed \cite{1} in the assumption of a background classical CS action with a 
purely time-like vector $\vec\eta = 0$ which, however,
is not consistent at the quantum level. The space-like scenario was recently
re-examined \cite{24,25} and the present situation can be conservatively characterized 
by an upper bound $|\vec\eta|\leq 10^{-32}\ \mbox{\rm eV}$.

For an arbitrary $\eta_\mu$, it is possible to realize that complex energy solutions 
arise iff $\eta^2 > 0$,
whilst real frequencies actually occur for any ${\vec k}$ if $\eta^2 < 0$, 
as it can be explicitly checked in the two cases described  
below\footnote{A general explicit and rigorous proof of this statement
is non-trivial and not yet reported in the Literature.}.

Let $\vec k$ orthogonal to $\vec\eta$, then the positive frequencies are given by
\begin{equation}
\omega_\perp^{(\pm)} (\vec k)=\sqrt{{\vec k}^2 +{1\over 2}{\vec\eta}^2 \pm \sqrt{{1\over 4}{\vec\eta}^4 +
{\vec k}^2 \eta_0^2}}\ .
\label{2.27}
\end{equation}

Furthermore, if $\vec k$ and  $\vec\eta$ are parallel and $\eta^2 < 0$, 
then the positive frequencies read
\begin{eqnarray}
\omega_{|\!|}^{(1)}(\vec k)=&&{1\over 2}|\vec\eta| + \sqrt{\vec k^2+{1\over 4}{\vec\eta}^2
-|\vec k| \eta_0 \mbox{\rm sgn}(\vec k\cdot\vec\eta)}\ ;\label{2.28a}\\
\omega_{|\!|}^{(2)}(\vec k)=&&{1\over 2}|\vec\eta| - \sqrt{{\vec k}^2+{1\over 4}{\vec\eta}^2
-|\vec k| |\eta_0|}\ \nonumber\\ 
&&\mbox{\rm iff}\ \ \eta_0 \vec k\cdot\vec\eta > 0\ 
\mbox{\rm and}\ \ |\vec k| < |\eta_0|\ ; \label{2.28b}\\
\omega_{|\!|}^{(3)}(\vec k)=&&- {1\over 2}|\vec\eta| + \sqrt{{\vec k}^2+{1\over 4}{\vec\eta}^2
+|\vec k| \eta_0 \mbox{\rm sgn}(\vec k\cdot\vec\eta)}\ \nonumber\\   
&& \mbox{\rm if}\ \mbox{\rm either}\ \ 
\eta_0 \vec k\cdot\vec\eta > 0\quad \mbox{\rm or}\ \ 
\eta_0\vec k\cdot\vec\eta < 0\ \mbox{\rm and}\ \ |\vec k| > |\eta_0|\ 
\ \  .\label{2.28c}
\end{eqnarray}
The frequency $\omega_\parallel^{(1)}(\vec k)$ characterizes the left-polarized photons whereas
the frequencies $\omega_\parallel^{(2),(3)}(\vec k)$ correspond to the right-polarized ones.
The two kinds of solutions (\ref{2.27}) and (\ref{2.28a})-(\ref{2.28c}) are separated by the light-cone ones: namely,
\begin{equation}
k^2_0 = \vec k^2,\ \ \vec k\cdot \vec\eta = \pm |\vec k| \eta_0\  . 
\label{2.29}
\end{equation}
This cone apart, the group velocities are less than one for $\eta^2 < 0$ and, consequently,
this pattern of Lorentz symmetry breaking appears to be suitable for field quantization \cite{21}.

Nonetheless, the solution (\ref{2.28a})-(\ref{2.28c}) indicates that, in general, the phenomenon of the 
frequency flip actually occurs. As a matter of fact,
for each direction $\vec k\cdot \vec\eta = |\vec k||\vec\eta|\cos\varphi,\ \ 
\varphi = \mbox{\rm constant}$, $\eta^2 < 0$, a zero-energy cone does indeed exist and reads
\begin{equation}
k_0 = 0,\ \ |\vec k| = \sqrt{\eta_0^2 - {\vec\eta}^2 \sin^2\varphi}.
\label{2.30}
\end{equation} 
Under small variations of the value of $|\vec k|$ in the vicinity of this cone,
it is possible to 
change smoothly the energy sign of distorted photons for the {\sl right} polarization (\ref{2.18}) and (\ref{2.22}), 
as it can be explicitly checked from eqs.~(\ref{2.28b}) and (\ref{2.28c}) for
different $\mbox{\rm sgn}(\eta_0 \vec k\cdot\vec\eta)$ when  $|\vec k|\simeq  |\eta_0|$.
In particular, for $\eta_0 \vec k\cdot\vec\eta > 0$ and $|\vec k| < |\eta_0|$ there are three positive
frequencies, one for left-polarized  photons and two for right-polarized photons. 
In the same case, the solution with a negative frequency exists only for left-polarized photons, 
whilst the situation becomes just the opposite for $\eta_0 \vec k\cdot\vec\eta < 0$ 
and $|\vec k| < |\eta_0|$. 

This frequency flip phenomenon makes the photon quantization ambiguous, because the specification of what 
a creation operator and what an annihilation operator are becomes momentum dependent. 
The only choice of $\eta_\mu$ which is free of the above ambiguity,
although essentially frame-dependent, is the purely space-like one
$\eta_\mu = (0, \vec\eta)$. 

In this particular case, let us obtain the free vector
propagator from the canonical quantization. To this aim, taking eqs.~(\ref{2.6}), (\ref{2.21}) 
and (\ref{2.22}) into account, the quantized gauge potential can be written
in the suitable form
\begin{eqnarray}
A_\nu(x)&&=\int{d{\vec k}\over (2\pi)^{3/2}\sqrt{2k_0}}
\left.
\varepsilon^{(L)}_\nu(k)\
a_L({\vec k})\exp\{-ik\cdot x\}\right|_{k_0=\omega_+({\vec k},{\vec\eta})}
+\ \mbox{\rm h.\ c.}\nonumber\\
&&+\int{d{\vec k}\over (2\pi)^{3/2}\sqrt{2k_0}}
\left.
\varepsilon^{(R)}_\nu(k)\
a_R({\vec k})\exp\{-ik\cdot x\}\right|_{k_0=\omega_-({\vec k},{\vec\eta})}
+\ \mbox{\rm h.\ c.}
\label{2.31}
\end{eqnarray}
where the creation and annihilation operators $a^\dagger_{L(R)}({\vec k}),
a_{L(R)}({\vec k})$ do fulfill the standard algebra. The free Feynman's
propagator of the Maxwell-Chern-Simons photon in the axial gauge can be readily derived
 from the above decomposition and reads
\begin{equation}
(-i)D_{\mu\nu}(k,\eta)={k^2 d_{\mu\nu}(k,\eta)\over 
(k^2+i\varepsilon)^2+\eta^2(k^2+i\varepsilon)-(\eta\cdot k)^2}\ ,
\label{2.32}
\end{equation}
where, for $\eta\cdot k\not=0$, the polarization tensor is given by
\begin{eqnarray}
d_{\mu\nu}(k,\eta)=&&-g_{\mu\nu}+{\eta_\mu k_\nu+\eta_\nu k_\mu\over \eta\cdot k}
\left(1+{\eta^2\over k^2}\right)
-{\eta_\mu\eta_\nu\over k^2}\nonumber\\
&&-\eta^2{k_\mu k_\nu\over (\eta\cdot k)^2}\left(1+{\eta^2\over k^2}\right)
-{i\over k^2}\epsilon_{\mu\nu\alpha\beta}\eta^\alpha k^\beta\ .
\label{2.33}
\end{eqnarray}
The above axial gauge propagator consistently fulfills the transversality
condition 
\begin{equation}
\eta^\mu D_{\mu\nu}(k,\eta)=0\ 
\label{2.34}
\end{equation} 
and turns out to be in agreement with
the expression reported in \cite{21}. Moreover, by its very construction from
the canonical quantization - see eq.~(\ref{2.31}) - it manifestly satisfies the 
requirement of microcausality. Nonetheless, one has to keep in mind that
perturbative calculations, beyond one loop, might generally lead to 
inconsistent results, owing to the presence of the so called double spurious
pole $(\eta\cdot k)^{-2}$ in the photon propagator. This is well known since
a long time, at least in the framework of perturbative  
Lorentz covariant non-abelian gauge theories \cite{22}.
\section{Free spinor field in the constant axial-vector background}
In this Section  
we shall discuss the main features concerning the propagation
in the {\it 3+1} dimensional Minkowski space-time 
of the free spinor field in the presence of a CPT and Lorentz covariance 
breaking kinetic term associated with a constant axial-vector $b_\mu$.
The free fermion spectrum can be obtained from the following modified 
Dirac's equation in the momentum representation: namely,
\begin{equation}
\left(\gamma^\mu p_\mu  -m -\gamma^\mu b_\mu\gamma_5\right)\psi = 0\ .
\label{3.1}
\end{equation}
After a straightforward algebra one finds that the free continuous spectrum is controlled by the on-shell condition
\begin{equation}
\left(p^2 + b^2 - m^2\right)^2 + 4 b^2 m^2 - 4 (b \cdot p)^2 = 0\ .
\label{3.2}
\end{equation}
Contrary to the boson case, 
this equation has real solutions for any value of $b_\mu$ .
However, the consistent quantization of the corresponding spinor field, 
in terms of the conventional anti-commuting creation and annihilation 
operators, actually can be performed
iff there are two pairs of opposite roots of eq.~(\ref{3.2})
and a mass gap between them. 
This condition holds true \cite{6} for sufficiently small $b_\mu$ and, 
in particular, the requirement of a mass gap between positive and negative 
frequencies can be obtained from the absence of solutions with $p_0 = 0$, 
which precisely  corresponds to ${\vec b}^2 < m^2$.

In the complementary range ${\vec b}^2 \geq m^2$ 
the frequency flip at $p_0 = 0$ occurs in two cases.
First, if in addition $b^2 \geq -  m^2$ then $p_0$ vanishes for
\begin{equation}
|\vec p| = \sqrt{|\vec b|^2 \cos^2\theta + b^2}  \pm \sqrt{|\vec b|^2 \cos^2\theta - m^2}. 
\label{3.3}
\end{equation}
where $\vec p\cdot \vec b \equiv |\vec p||\vec b|\cos\theta$.  
Second, if  $b^2 \leq -  m^2$, the frequency flip occurs for
\begin{equation}
|\vec p| =\sqrt{|\vec b|^2 \cos^2\theta - m^2} \pm 
\sqrt{|\vec b|^2 \cos^2\theta + b^2} . 
\label{3.4}
\end{equation}  
Both solutions are derived at a given direction inside the cone
$\cos^2\theta \geq m^2/|\vec b|^2 $ which overlaps, in the particular case
$b^2 <0$, with a further cone provided by the
additional bound $\sin^2\theta \leq b_0^2/|\vec b|^2$.
In the general situation $b_0 \vec b\cdot\vec p\not=0$, one has  only simple
zeroes at $p_0 = 0$ and, therefore, in the vicinity of the cones (\ref{3.3}) and 
(\ref{3.4}) the two pairs of opposite frequencies are converted into three positive 
and one negative frequencies or {\it vice versa}.

The existence of the mass gap is not the only requirement to perform
the consistent quantization \cite{6}: one should also ask for group
velocities to be always less than or equal to the speed of light.
This holds true in the time-like case $b^2 > 0$, as it can be explicitly 
checked in four particular solutions displaying the behavior of energies 
in different regions of the $b$-parameter space.

For a purely time-like $b^\mu = (b_0,0,0,0)$, one easily finds that
\begin{equation}
p_0^2 = \left(|\vec p|\pm b_0\right)^2 + m^2\ ,
\label{3.5}
\end{equation}
where the positive and negative  
signs just correspond to  positive and negative 
fermion helicities respectively. 
This dispersion law unravels a different kinematics for 
low and high momenta. Namely, for $|\vec p| \leq ( b_0^2 + m^2)/ 2|b_0|$ 
both types of solutions behave like massive states with $p^2 > 0$, 
whereas for higher momenta $|\vec p| \geq ( b_0^2 + m^2)/ 2|b_0|$ the solutions
with a negative helicity have $p^2 < 0 $, 
which eventually leads to instability of high-energy fermions and their
decay into pairs of fermions and antifermions \cite{6}, once interaction with photons
has been switched on (see Section 4).

Let us now turn our attention to the purely space-like case $b_\mu = (0,\vec b)$,
where the dispersion law is given by
\begin{equation}
p_0^2 = \vec p^2 +\vec b^2 + m^2  
\pm 2|\vec b|\sqrt{\vec p^2 \cos^2\theta + m^2}\ .
\label{3.6}
\end{equation}
These solutions are also separated by the stability cone.
For arbitrary $b_\mu$ the stability border $p^2 = 0$ is described by
\begin{equation}
|p_0| = |\vec p| = {|b_0^2 -\vec b^2 + m^2| \over 2 |b_0 \mbox{\rm sgn}(p_0)
 - |\vec b|\cos\theta|}\ .
\label{3.7}
\end{equation}
Two more solutions can be presented in a simple form when 
$\theta = n\pi/2,\ n\in{\bf Z}$.
In the orthogonal case $n=2k+1,\ k\in{\bf Z}$, they read
\begin{equation}
p_0^2 = \vec p^2 + b_0^2 +\vec b^2 + m^2  
\pm 2\sqrt{ b_0^2 \vec p^2 +\vec b^2( b_0^2 + m^2)} > 0\ ,
\label{3.8}
\end{equation}
corresponding to two pairs of opposite frequency solutions.
In the parallel case $n=2k,\ k\in{\bf Z}$, the positive and negative 
frequencies  are no longer opposite pairs because
\begin{eqnarray}
p_{0\pm}^{(1)} =&&\ |\vec b|   \pm 2\sqrt{ [|\vec p| - b_0 
\mbox{\rm sgn}(\vec p\cdot\vec b)]^2 + m^2}\ ,\label{3.9a}\\
p_{0\pm}^{(2)} =&& - |\vec b|   \pm 2\sqrt{ [|\vec p| + b_0 
\mbox{\rm sgn}(\vec p\cdot\vec b)]^2 + m^2}\ . \label{3.9b}
\end{eqnarray}
The explicit solutions presented above are such that the corresponding
group velocities are less than one iff $b^2 > 0$. So we can conclude that
the free field theory of fermions 
satisfying the modified Dirac equation (\ref{3.1})
is perfectly consistent and causal in 
the time-like case $b^2 > 0$, even if space-like
1-particle state appear.

Let us finally analyze the physical effects of Lorentz symmetry breaking
due to the presence of a constant axial-vector $b_\mu$
in the free spinor kinematics. 
We suppose it to be essentially time-like, $b_0^2 > \vec b^2$. 
Taking the relatively small momenta $|\vec p| \leq m$, one finds 
the effective mass splitting 
between energies of fermions of different helicities 
which is basically controlled by the time component $b_0$. 
In particular, for the purely time-like $b_\mu = (b_0,0,0,0)$, the relative
energy splitting is approximately given by
\begin{equation}
{\Delta p_0 \over p_0} \simeq {4|\vec p| b_0 \over \vec p^2 + m^2};\ \
\mbox{\rm max }\ {\Delta p_0 \over p_0} \simeq {2b_0 \over m}\ \ \mbox{\rm for}\ \  |\vec p|\simeq m .
\label{3.10}
\end{equation}
Evidently, if the value of $b_0$ is universal for all fermions, 
then the lightest massive ones - electrons
and positrons -  should give the best precision in its determination.
As the precision of a measurement of the electron mass   
\cite{26} is of the order $10^{-8}$, then the upper bound on the 
time component of CPT breaking 
vector is not very stringent: namely, $|b_0| < 10^{-2}\ \mbox{\rm eV}$ 
(or even weaker, being controlled by the energy 
resolution in accelerator beams).
On the other hand,  much more stringent
bounds were obtained for space components of $b_\mu$, 
from the experiments with atomic systems using 
hydrogen masers \cite{27} - {\it i.e.} $|\vec b| < 10^{-18}\ \mbox{\rm eV}$ - 
and with a spin-polarized torsion pendulum \cite{28} - {\it i.e.}
$|\vec b| < 10^{-20}\ \mbox{\rm eV}$. 
Thus the present bounds leave a room for a tiny, time-like
CPT breaking due to the possible presence of a constant axial-vector $b_\mu$.

A further phenomenon would take place at very 
high energies, $M > m^2/|b_0| > 10^2\ \mbox{\rm TeV}$ -
which represents the second, ultraviolet scale
of this kind of QED - when some of the electron states
reach the space-like four-momenta. Then, the conventional wisdom of the Special
Relativity Theory would fail and a highly energetic electron 
might well decay into an electron 
and a pair of positron and electron, as 
we shall further detail in the next Section.
\section{Physical cutoff for fermions and radiatively induced CS vertex}
As it was explained in the previous Section, one type of fermions
achieves the space-like four-momentum $p^2 < 0$ at very high energies, 
a special phenomenon which would break the conventional
Lorentz kinematics in the scattering and decay processes. As a consequence, 
high-energy fermions of a given polarization turn out to become unstable
once their interaction with photons is switched on. 
Let us consider the electrons of a given helicity and
four-momentum $p_\mu$ and 
analyze the possibility for them to
decay into an electron of the same helicity and with momentum 
$k^{(1)}_\mu$, together with a couple of
electron and positron of momenta $k^{(2)}_\mu$ and $k^{(3)}_\mu$ 
carrying the negative and positive helicities respectively -
keeping in mind the helicity conservation at high energies.
For simplicity but without loss of generality, we restrict ourselves 
to the purely time-like $b_\mu= (b_0,0,0,0)$ and take $b_0 > 0$. 

In order to describe the kinematics near the threshold of such a reaction,
one can parameterize the space momenta near the forward direction
\begin{eqnarray}
&&\vec k^{(j)} = \beta_j \vec p + \vec\Delta_j\ ;\quad 
 \vec p\cdot \vec\Delta_j = 0\ ; \quad
|\vec\Delta_j| \ll |\vec p|\ ;\nonumber\\
&&\sum_{j=1}^3 \beta_j = 1\ ;\quad \sum_{j=1}^3\vec\Delta_j = 0\ ,
\label{4.1}
\end{eqnarray}
where the latter equations follow from the momentum conservation.
The energy conservation in this case  yields the following relationship:
namely,
\begin{eqnarray}
&&\sqrt{\left(|\vec p|\pm b_0\right)^2 + m^2} =\nonumber\\
&&\sqrt{\left(|\vec  k^{(1)}|\pm b_0\right)^2 + m^2}+ 
\sqrt{\left(|\vec  k^{(2)}| - b_0\right)^2 + m^2} + \sqrt{\left(|\vec  
k^{(3)}|- b_0\right)^2 + m^2}\ ; \label{4.2a}\\
&&|\vec p| \pm b_0 + {m^2 \over 2|\vec p|}\ \simeq\ |\vec p|\sum_{j=1}^3
\beta_j \pm b_0 - 2 b_0 + \sum_{j=1}^3
{m^2 + \vec\Delta_j^2 \over 2\beta_j|\vec p|}\ ,
\label{4.2b}
\end{eqnarray}
where the dispersion law (\ref{3.5}) has been used. Evidently, in order the decay 
process could start with, the fermion has to reach the momentum
\begin{equation}
|\vec p| \simeq 
{1 \over 4 b_0}\left\{\sum_{j=1}^3 
{m^2 + \vec\Delta_j^2 \over \beta_j}  - m^2\right\}\ .
\label{4.3}
\end{equation}
Therefrom, the minimal value of $|\vec p|$ 
is achieved when $\beta_j = 1/3$ and  $\vec\Delta_j^2 = 0$: namely, 
\begin{equation}
\mbox{\rm min }|\vec p| \simeq {2 m^2 \over b_0} \equiv \Lambda_s.
\label{4.4}
\end{equation}
We remark that this  lower bound endorses the profile of the stability bound 
of eq.~(\ref{3.7}) and is
isotropic for $b^\mu= (b_0,0,0,0)$. 

Thus one concludes that, once the interaction with photons
is turned on, the high-energy fermion states of a definite
polarization become unstable and, strictly speaking, the
corresponding 1-particle states can not appear as asymptotic 
incoming and outgoing states of the physical Hilbert space.
In such a modified QED the {\sl physical cutoff} in the three-momentum space
necessarily arises for the intermediate 1-particle physical states,
at least in a non-perturbative approach. 
Here we would like to focus on perturbation theory, {\it i.e.} on 
the one-loop radiatively induced CS vertex for photons due to the 
presence of the constant axial-vector $b^\mu= (b_0,0,0,0)$. 
Let us derive the relevant coefficient to the
first order in  $b_0$, as the value of that coefficient appears to be
quite controversial in the recent Literature \cite{3},\cite{13}-\cite{20}.

Our aim is to compute the one-loop induced CS parity-odd effective 
action from the classical spinor Lagrange density  
\begin{equation}
{\cal L}_{\mbox{\rm spinor}}=\bar\psi\left(i\gamma^\mu\partial_\mu
+e\gamma^\mu A_\mu -m -\gamma^\mu b_\mu\gamma_5\right)\psi\ ,
\label{4.5}
\end{equation}
which leads to the momentum space four-dimensional Feynman's propagator
\begin{equation}
(-i)S(p)={p^2+ b^2-m^2+2\left(b\cdot
p+m b_\mu\gamma^\mu\right)\gamma_5\over
\left(p^2+b^2-m^2+i\varepsilon\right)^2-4\left[(b\cdot
p)^2-m^2 b^2\right]}\left( \gamma^\nu p_\nu
+m+b_\nu\gamma^\nu\gamma_5\right)\ . 
\label{4.6}
\end{equation}
>From the Feynman's rules, the one-loop photon self-energy tensor 
is formally determined to be
\begin{equation}
\Pi^{\mu\nu}(k)=\int{d^{4}p\over
i(2\pi)^{4}}\ {\tt tr}\left\{\gamma^\mu S(p)\gamma^\nu
S(p-k)\right\}\ .
\label{4.7}
\end{equation}
However, the above formal expression exhibits by
superficial power counting ultraviolet divergences 
which have to be properly regularized.
The general structure
of the regularized photon self-energy tensor turns out to be
\begin{equation}
\mbox{\tt reg}\Pi^{\mu\nu}(k)={\tt reg}\Pi^{\mu\nu}_{\mbox{\rm even}}(k)
+{\tt reg}\Pi^{\mu\nu}_{\mbox{\rm odd}}(k)\ .
\label{4.8}
\end{equation}
The parity-even part will be considered elsewhere. Here
our goal is to derive the radiatively induced CS constant for the
parity-odd part.

In order to derive the analytic expression for the parity-odd part 
one first performs  
the trace at the numerator in eq.~(\ref{4.6}), which will be done in details
in the next Section - see eq.~(\ref{5.3}) and further on.
Then the relevant part of the integral, 
to the lowest order in $b_\mu$ and in the
external momentum, can be cast in the following form, which manifestly
fulfills the invariance under the maximal residual symmetry subgroup
$SO(3)$, namely:
\begin{eqnarray}
{\tt reg}\Pi^{\mu\nu}_{\mbox{\rm odd}}(k;b,m)&&\simeq -4
\epsilon^{\mu\nu\rho\sigma} k_\sigma
\int_{|\vec p|\leq \Lambda_s }{{d^3 p}\over
(2\pi)^{4}}\int_{-\infty}^{+\infty} dp_0\ {b_\rho 
(p^2 + 3m^2)- 4 p_\rho (b\cdot p) \over (p^2-m^2+i\varepsilon)^3}\nonumber\\
&&= -4
\epsilon^{\mu\nu 0\sigma} b_0 k_\sigma
\int_{|\vec p|\leq \Lambda_s }{{d^3 p}\over
(2\pi)^{4}}\int_{-\infty}^{+\infty}
 dp_0\ { 3m^2 - 3p_0^2 - \vec p^2 \over (p^2-m^2+i\varepsilon)^3}\nonumber\\
&&= 
\epsilon^{\mu\nu 0\sigma} b_0 k_\sigma 
{i \over 2\pi^2}\left(1 + { m^2\over \Lambda^2_s}\right)^{- 3/2}
\simeq \epsilon^{\mu\nu 0\sigma} b_0 k_\sigma 
{i \over 2\pi^2}\ .
\label{4.9}
\end{eqnarray}
Here a large 3-momentum cutoff $\Lambda_s$ has been introduced to remind us
the unavoidable presence of the stability bound (\ref{4.4}) at the non-perturbative
level.
The integral over $p_0$ is convergent and can be easily calculated 
from the residues theorem by closing, for instance, the contour in 
the upper complex plane. 
Thus the induced CS term comes out solely from the mass-shell contribution 
and this is why it is purely imaginary. 
The remaining integral over $\vec p$
is a standard one and altogether we definitely find that
the induced CS term is not vanishing and its value is uniquely
accounted for by the requirement of the manifest invariance under
the maximal residual symmetry subgroup $SO(3)$, a requirement which
definitely amounts to be the physical renormalization prescription.

On the other hand one can see that, in
fact, the value of the large physical cutoff $\Lambda_s$
does not make any influence on the leading, mass-independent CS term,
as it does because we are dealing with a perturbative calculation. 
The basic ingredient to obtain the result of eq.~(\ref{4.9}) 
is nothing but the integration prescription.
First one should perform the integration over the small group of 
$b_\mu$ which is the $SO(3)$ rotation group for time-like $b^\mu$.
In particular, for purely time-like $b_\mu$ 
the average over rotations of the spatial  3-momentum  yields
$\left< 4 p_\rho (b\cdot p)\right>_{D=3} = 4 b_0 p^2_0\delta_{\rho 0}$, 
the result  which 
has been used in eq.~(\ref{4.9}).  We emphasize this point 
to explain the discrepancy which arises between
different regularization approaches. 
In Refs.~\cite{13,14} the assumption of the exact Lorentz
symmetry in the 4-momentum space was adopted, {\it i.e.}, 
the averaging relation
$\left<4 p_\rho (b\cdot p)\right>_{D=4} = b_\rho p^2$ was employed. 
However, the Lorentz symmetry is certainly broken and only the rotational 
3-momentum symmetry survives in the exact expression (\ref{4.7}) for 
the polarization tensor when $b^\mu$ is time-like.

Next step is to replace the denominator under
integration by the mass-shell $\delta$ distribution
\begin{equation}
{1 \over (p^2-m^2+i\varepsilon)^3} \rightarrow -{i\pi\over 2} 
\delta''(p^2 - m^2),
\label{4.10}
\end{equation}
as the final result is purely imaginary. One can easily prove that
in this way only finite integrals appear in eq.~(\ref{4.9}) yielding eventually the
unique result for the induced CS coupling constant.

Finally, we remark that for arbitrary time-like $b^\mu$
the threshold momentum min$|\vec p|$ for the decay will depend 
on the direction, {\it i.e.} on $\cos\theta$ in a certain 
correspondence to eq.~(\ref{3.7}). 
Indeed, the kinematical bounds are  determined by the mass-shell
equation (\ref{3.2}) which, in turn, is invariant under simultaneous Lorentz
transformation $p'_\mu = L^{\nu}_\mu p_\nu$ and $b'_\mu = L^{\nu}_\mu b_\nu$.
Therefore the threshold position will be transformed accordingly,
the isotropic cutoff being converted into an anisotropic one, while keeping
the result of integration in eq.~(\ref{4.9}) invariant.

Further on, we will not develop the present physical approach 
to calculate the full one-loop induced effective action.
Instead of, we propose in the next Section the equivalent formalism of 
dimensional regularization which reproduces the same CS vertex, 
after being adapted to the Lorentz symmetry breaking.
\section{Induced Chern-Simons term in the 
dimensional regularization}
Let us now calculate the induced parity-odd  effective 
action by means of dimensional regularization properly adjusted
to the Lorentz symmetry breaking phenomenon.
We start from the same spinor Lagrange density (\ref{4.5}) and the related
Feynman's propagator (\ref{4.6}) both defined strictly in four dimensions.

Notice that, at least in the special cases of a purely time-like 
and/or purely space-like four-vector $b_\mu$, the four-dimensional 
free fermion propagator has two pairs of opposite simple poles 
on the real $p_0$ axis which are regulated according to Feynman's causal prescription. 
We also stress that the free fermion propagator  
is obtained after inversion of the kinetic operator 
appearing in the classical Lagrange density (\ref{4.5}) in terms of 
the conventional four-dimensional Clifford's algebra of the Dirac's matrices.

The one-loop photon self-energy tensor is thereby given
\begin{equation}
{\tt reg}\Pi^{\mu\nu}(k)=\mu^{4-2\omega}\int{d^{2\omega}p\over
i(2\pi)^{2\omega}}\ {\tt tr}\left\{\gamma^\mu S(p)\gamma^\nu
S(p-k)\right\}\ ,
\label{5.1}
\end{equation} 
where dimensional regularization is employed to give a meaning to the loop integral, 
which appear by power counting to be superficially quadratically divergent in four dimensions. 
Notice that the whole set of the Dirac's matrices involved in the regularized loop integral (\ref{5.1}) 
has now to be understood and treated according to the algebraically consistent 
general rules suggested by 't Hooft-Veltman-Breitenlohner-Maison \cite{29}.  
The trace at the numerator in eq.~(\ref{4.3}) amounts to be 
\begin{eqnarray}
&&{\tt tr}\left\{\bar\gamma^\mu\left(p^2+b^2-m^2+2 b\cdot
p\gamma_5+ 2m b_\alpha\bar\gamma^\alpha\gamma_5\right)\left(
\gamma^\beta p_\beta + m
+ b_\beta\bar\gamma^\beta\gamma_5\right)\bar\gamma^\nu\right.\nonumber\\
&&\left.\left[(p-k)^2+ b^2-m^2+2 b\cdot (p-k)\gamma_5+
2m b_\lambda\bar\gamma^\lambda\gamma_5\right]\left[
\gamma^\sigma (p-k)_\sigma + m
+ b_\sigma\bar\gamma^\sigma\gamma_5\right]\right\}\ ,
\label{5.2}
\end{eqnarray}
where we have taken into account that the external indices $\mu,\nu$ as well as the four-vector $b_\alpha$ are
physical, {\it i.e.} $\mu,\nu,\alpha=0,1,2,3$ so that, consequently, the corresponding matrices $\bar\gamma^\mu,\bar\gamma^\nu$ are physical 
and contraction of $b_\alpha$ with a $\gamma$-matrix always involves a 
$\bar\gamma^\alpha$ matrix - see \cite{29} and Appendix A.  The general structure
of the photon self-energy tensor is again presented by (\ref{4.8}).

Here we are interested in the CS parity-odd term: the only non-vanishing
contributions to such a term are given by the traces of the products of six
$\gamma$-matrices with $\gamma_5$ and of four $\gamma$-matrices with
$\gamma_5$, since the traces of the products of two, three and five
$\gamma$-matrices with $\gamma_5$ do indeed vanish in $2\omega$-dimensions. 
A straightforward computation gives
\medskip
\noindent
(a)\ six $\gamma$-matrices and $\gamma_5$: 
\begin{eqnarray}
&&4m^2{\tt
tr}\left\{\bar\gamma^\mu b_\alpha\bar\gamma^\alpha\gamma_5\left(\gamma^\beta
p_\beta+ b_\beta\bar\gamma^\beta\gamma_5\right)\bar\gamma^\nu
b_\lambda\bar\gamma^\lambda\gamma_5\left[\gamma^\sigma
(p-k)_\sigma+ b_\sigma\bar\gamma^\sigma\gamma_5\right]\right\}\nonumber\\
&&=-4m^2 b^2(i2^\omega)\epsilon^{\mu\nu\rho\sigma} b_\rho k_\sigma +
\mbox{\rm even\ terms}\ ;
\label{5.3}
\end{eqnarray}
\medskip
\noindent
(b)\ four $\gamma$-matrices and $\gamma_5$: the parity-odd contribution are of
two kinds 
\begin{equation}
Q^{\mu\nu}=-A_1(i2^\omega) \epsilon^{\mu\nu\rho\sigma} b_\rho k_\sigma 
           +A_2(i2^\omega) \epsilon^{\mu\nu\rho\sigma}p_\rho k_\sigma\ ,
\label{5.4}
\end{equation}
with
\begin{eqnarray}
A_1&&=2m^2\left(p^2+ b^2-m^2\right)
   +2m^2\left[(p-k)^2+ b^2-m^2\right]\nonumber\\
   &&+\left(p^2+ b^2-m^2\right)\left[(p-k)^2+ b^2-m^2\right]
    +4(b\cdot p)[b\cdot (p-k)]\ ,
\label{5.5}
\end{eqnarray}
\begin{equation}
A_2=2(b\cdot p)\left[(p-k)^2+b^2-m^2\right]
   +2[b\cdot (p-k)]\left(p^2+b^2-m^2\right)\ .
\label{5.6}
\end{equation}
Putting all together we find that the one-loop CS parity-odd part of the
photon self-energy tensor takes the form
\begin{eqnarray}
&&{\tt reg}\Pi^{\mu\nu}_{\mbox{\rm odd}}(k;b,m)=-2^\omega
\mu^{4-2\omega}\epsilon^{\mu\nu\rho\sigma} k_\sigma
\int{d^{2\omega}p\over
(2\pi)^{2\omega}}\nonumber\\
&&\left\{b_\rho F^{(1)}(p;k,b,m)-p_\rho F^{(2)}(p;k,b,m)\right\}
D(p;b,m)D(p-k;b,m)\ ,
\label{5.7}
\end{eqnarray}
where
\begin{eqnarray}
F^{(1)}(p;k,b,m)&&=2m^2\left[p^2+(p-k)^2 -2m^2\right]+4(b\cdot p)b\cdot (p-k)
\nonumber\\
&&+
\left(p^2+ b^2-m^2\right)\left[(p-k)^2+b^2-m^2\right]\ ,
\label{5.8}
\end{eqnarray}
\begin{equation}
F^{(2)}(p;k,b,m)=
2(b\cdot p)\left[(p-k)^2+b^2-m^2\right]
+2[b\cdot (p-k)]\left(p^2+b^2-m^2\right)\ ,
\label{5.9}
\end{equation}
whereas the scalar propagator reads
\begin{equation}
D(p;b,m)=\left\{(p^2+b^2-m^2+i\varepsilon)^2-4[(b\cdot p)^2-b^2
m^2]\right\}^{-1}\ .
\label{5.10}
\end{equation}

Let us now consider the integral involving $F^{(2)}(p;k,b,m)$:
\begin{eqnarray}
\int{d^{2\omega}p\over
(2\pi)^{2\omega}}\ &&p_\rho F^{(2)}(p;k,b,m)D(p;b,m)D(p-k;b,m)=\nonumber\\
&&b_\rho I^{(2)}(k;b,m)+ k_\rho J^{(2)}(k;b,m)\ ;
\label{5.11}
\end{eqnarray}
the contribution involving $J^{(2)}(k;b,m)$ is clearly irrelevant to our
aim, whereas that one involving $I^{(2)}(k;b,m)$ can be isolated after
contraction with the four-vector 
\begin{equation}
\tilde b^\rho={k^2 b^\rho -(b\cdot k)k^\rho\over 
b^2 k^2 -(b\cdot k)^2}\ ,\quad
\tilde b^\rho k_\rho=0\ ,\quad \tilde b^\rho b_\rho=1\ .
\label{5.12} 
\end{equation}
The result is
\begin{equation}
{\tt reg}\Pi^{\mu\nu}_{\mbox{\rm odd}}(k;b,m)=2^\omega
\mu^{4-2\omega}\epsilon^{\mu\nu\rho\sigma}b_{\rho} k_\sigma\
{\tt reg}\Pi_{\mbox{\rm odd}}(k;b,m)\ ,
\label{5.13}
\end{equation}
where
\begin{eqnarray}
&&{\tt reg}\Pi_{\mbox{\rm odd}}(k;b,m)=\nonumber\\
&&\int{d^{2\omega}p\over
(2\pi)^{2\omega}}
\left\{G^{(2)}(p;k,b,m)-F^{(1)}(p;k,b,m)\right\}
D(p;b,m)D(p-k;b,m)\ ,   
\label{5.14}
\end{eqnarray}
with
\begin{equation}
G^{(2)}(p;k,b,m)\equiv F^{(2)}(p;k,b,m)\
{k^2 b\cdot p -(b\cdot k)(k\cdot p)\over
b^2 k^2 -(b\cdot k)^2}\ .
\label{5.15}
\end{equation}

Now, owing to the Feynman's causal prescription to regulate poles, we can
compute the integral (\ref{5.14}) under transition to the $2\omega$-dimensional
Euclidean space, {\it i.e.}, under the replacements
\begin{eqnarray}
&&p_0=ip_4\ ,\quad p^2=-p_E^2\ ,\quad
b_0=ib_4\ ,\quad b^2=-b_E^2\ ,\quad
k_0=ik_4\ ,\quad k^2=-k_E^2\ ;\nonumber\\
&&p_{E\mu}=(p_E,\phi,\theta_1,\ldots,\theta_{2\omega-3},\theta)\ ,\quad
d^{2\omega}p=id^{2\omega}p_E\ ,\quad q=p_E^2\ ,\nonumber\\
&&d^{2\omega}p_E={1\over 2}
q^{\omega-1}dq d\phi(\sin\theta)^{2\omega-2}d\theta
\prod_{n=1}^{2\omega-3}\sin^n\theta_n d\theta_n\ ;\label{5.16}\\
&&b\cdot p=-b_E p_E\cos\theta\ ,\quad b\cdot k=-b_E
k_E\cos\Theta\ ,\quad
k\cdot p=-k_E p_E(\cos\theta_1+\cos\theta\cos\Theta)\ ;\nonumber\\
&&0<q<\infty\ ,\quad 0<\phi<2\pi\ ,\quad 0<\theta<\pi\ ,\quad
0<\theta_n<\pi\ .\nonumber
\end{eqnarray}

The denominator in eq.~(\ref{5.14}) then becomes the product of the Euclidean
propagators 
\begin{eqnarray}
&&D_E(p_E;b_E,m)=\left\{(
p_E^2+m^2-b_E^2)^2+4 b_E^2 p_E^2\sin^2\theta\right\}^{-1}\ ,\label{5.17a}\\
&&D_E(p_E-k_E;b_E,m)=\left\{[p_E^2+k_E^2-2 p_E
k_E(\cos\theta_1+\cos\theta\cos\Theta)+m^2-b_E^2]^2\right.\nonumber\\
&&\left.+4b_E^2(p_E^2\sin^2\theta+k_E^2\sin^2\Theta-2 p_E
k_E\cos\theta_1)\right\}^{-1}\ ,\label{5.17b}
\end{eqnarray}
whilst the numerator can be rewritten as a sum of two contributions: namely,
\begin{eqnarray}
&&N_E^{(a)}(p_E;k_E,b_E,m)=F^{(1)}(p;k, b,m)+{(b\cdot
p)k^2\over b^2k^2-(b\cdot k)^2} F^{(2)}(p;k,b,m)=\nonumber\\
&&-2m^2[p_E^2+(p_E-k_E)^2+2m^2]
+(p_E^2+b_E^2+m^2)[(p_E-k_E)^2+b_E^2+m^2]\nonumber\\
&&+4 b_E^2p_E^2\cos^2\theta -4b_E^2 p_E
k_E\cos\theta\cos\Theta\nonumber\\
&&-2 p_E^2\cos^2\theta\csc^2\Theta[p_E^2+(p_E-
k_E)^2+2b_E^2+2m^2]\nonumber\\
&&+2 p_E k_E\cos\theta\cos\Theta\csc^2\Theta(p_E^2+b_E^2+m^2)\ ;
\label{5.18a}\\ 
&&N_E^{(b)}(p_E;k_E,b_E,m)={(b\cdot k)(p\cdot k)\over
b^2k^2-(b\cdot k)^2} F^{(2)}(p;k,b,m)=\nonumber\\
&&-2p_E^2\cos\theta(\csc^2\Theta\cos\Theta\cos\theta_1+\cot^2\Theta\cos\theta)
[(p_E-k_E)^2+ p_E^2 +2b_E^2 +2m^2]\nonumber\\
&&+2 p_E k_E\cot^2\Theta(\cos\theta_1+\cos\theta\cos\Theta)
(p_E^2+b_E^2+m^2)\ . \label{5.18b}
\end{eqnarray}

The coefficient of the radiatively induced CS parity-odd Lagrange density can
be obtained from the above quantities in the limit $k_E\to 0$: then we find
\begin{eqnarray}
&&N_E^{(a)}(p_E;0,b_E,m)-N_E^{(b)}(p_E;0,b_E,m)\nonumber\\
&&=(p_E^2+m^2-b_E^2)^2+4 p_E^2b_E^2\sin^2\theta\nonumber\\
&&-4(p_E^2+m^2-b_E^2)(m^2+ p_E^2\cos^2\theta)\nonumber\\
&&+4 p_E^2(p_E^2+m^2+b_E^2)\cos\theta\cos\theta_1\cos\Theta
\csc^2\Theta\ .
\label{5.19}
\end{eqnarray}
and it follows therefrom that we have to evaluate the Euclidean quantity
\begin{eqnarray}
&&{\tt reg}\Pi_{E,\mbox{\rm odd}}(0;b_E,m)=\nonumber\\
&&\int{d^{2\omega}p_E\over
i(2\pi)^{2\omega}}
\left\{N_E^{(a)}(p_E;0,b_E,m)-
N_E^{(b)}(p_E;0,b_E,m)\right\} \left[
D_E(p_E;b_E,m)\right]^2\ .
\label{5.20}
\end{eqnarray}
Integration with respect to the angular variables is straightforward and
yields \cite{30} 
\begin{eqnarray}
\Pi_{E,\mbox{\rm odd}}(0;b_E,m)&&\equiv \lim_{\omega\uparrow 2}{\tt reg}\Pi_{E,\mbox{\rm odd}}(0;b_E,m)\nonumber\\
&&={(-i)\over 16\pi^2}\int_0^\infty dq\ q
\left\{{1\over |z(q)|\sqrt{A(q)}}-4m^2\mbox{\rm sgn}[z(q)][A(q)]^{-3/2}\right.\nonumber\\
&&-\left.{4[z(q)+b_E^2]^2-4m^2[z(q)+b_E^2]\over
|z(q)|\sqrt{A(q)}[|z(q)+\sqrt{A(q)}]^2}\right\}\nonumber\\
&&+\lim_{\omega\uparrow 2}\ {i(2-\omega)\over 
4\omega\pi^{\omega}\sqrt\pi}{\Gamma\left(\omega-1/2\right)\over
\Gamma\left(2\omega-1\right)}\int_0^\infty dq\ {q^\omega\over (q+m^2)^3}
\ ,
\label{5.21}
\end{eqnarray}
where we have set 
\begin{equation}
z(q)\equiv q+m^2-b_E^2\ ,\qquad A(q)\equiv [z(q)]^2+4q b_E^2\ .
\label{5.22}
\end{equation}
We notice that the last term in the RHS of eq.~(\ref{5.19}) does not contribute since
the integration with respect to the variable $\theta_1$ obviously vanishes.
A further important remark is also in order. 
In the first integral of the RHS of eq.~(\ref{5.21}) we have definitely set $\omega=2$, because
there is no divergent part  as the integrand as a
whole is ${\cal O}(q^{-2})$ for large $q$. On the contrary, it is not possible to set
$\omega=2$ in the very last term of the above eq.~(\ref{5.21}), as it does involve the product of
the vanishing factor $(2-\omega)$ times a divergent integral when $\omega=2$.
Consequently, we have to take the limit $\omega\uparrow 2$ after the integration -
{\it i.e.} in the sense of the distributions - and the result is
\begin{equation}
\lim_{\omega\uparrow 2}\ {i(2-\omega)\over 
4\omega\pi^{\omega}\sqrt\pi}{\Gamma\left(\omega-1/2\right)\over
\Gamma\left(2\omega-1\right)}\int_0^\infty dq\ {q^\omega\over (q+m^2)^3}
={i\over 32\pi^2}\ .
\label{5.23}
\end{equation}
Now, the first integral of the RHS of eq.~(\ref{5.21}) can be easily calculated
\cite{30} for arbitrary values of $b_E^2\not=m^2$ and drives precisely to
the result quoted in Ref.~\cite{14}. On the other hand the additional contribution (\ref{5.23}), 
whose presence is unavoidable since the integral of eq.~(\ref{5.20}) does not exist in
four dimensions, is such that the final result reads 
\begin{equation}
\Pi_{E,\mbox{\rm odd}}(0;m^2/b_E^2)=
{i\over 8\pi^2}\left\{\vartheta(m^2-b_E^2)
+\vartheta(b_E^2-m^2)\left(1-\sqrt{1-{m^2\over b_E^2}}\right)\right\}\ , 
\label{5.24}
\end{equation}
where $\vartheta$ is the Heaviside's step distribution.

Under transition back to the Minkowski space-time we obtain
the radiatively induced Chern-Simons CPT-odd term, within
dimensional regularization: namely,
\begin{eqnarray}
\Pi^{\mu\nu}_{\mbox{\rm odd}}(0;b,m)&&\equiv\lim_{\omega\to 2}
{\tt reg}\Pi^{\mu\nu}_{\mbox{\rm odd}}(k=0;b,m)
=4\epsilon^{\mu\nu\rho\sigma} b_\rho k_\sigma\Pi_{\mbox{\rm odd}}(k=0;m^2/ b^2)\nonumber\\
&&={i\over 2\pi^2}\epsilon^{\mu\nu\rho\sigma} b_\rho k_\sigma
\left\{1-\vartheta(-b^2-m^2)\sqrt{1+{m^2\over b^2}}
\right\}\ .
\label{5.25}
\end{eqnarray}
 We definitely find that
 the induced CS term of eq.~(\ref{5.25}) {\sl exactly} coincides with 
the CPT-odd term (\ref{4.9}) derived with the help of the physical cutoff.

Let us now outline the roots of recent controversies in the calculation of the
latter CS constant. 
To this concern, it is well known since a long
time \cite{31}  - and also recently reconsidered  within the present context \cite{32} -
that the {\sl perturbative} self-energy tensor is defined up to a prescription
dependent boundary term in such a way that we can
write in general, 
\begin{equation}
\Pi^{\mu\nu}_{\mbox{\rm odd}}[\zeta]=\Pi^{\mu\nu}_{\mbox{\rm odd}}
-{i\zeta\over 2\pi^2}\epsilon^{\mu\nu\rho\sigma} b_\rho k_\sigma\ ,\quad
\zeta\in{\bf R} 
\label{5.26}
\end{equation}
consistently with the Ward's identity $k_\mu\Pi^{\mu\nu}_{\mbox{\rm odd}}[\zeta]=0$.
In other words, different renormalization 
prescriptions lead to expressions of $\Pi^{\mu\nu}_{\mbox{\rm odd}}$ which are always
finite but different from our eq.~(\ref{5.25}) up to the local polynomial
$(i\zeta/2\pi^2)\epsilon^{\mu\nu\rho\sigma} b_\rho k_\sigma$,
the only
one allowed by power counting and symmetry properties. 

In the
configuration space we actually obtain a one-parameter family of 
the one-loop radiatively induced CPT-odd Lagrange densities, {\it i.e.}, 
\begin{equation}
{\cal L}_{\mbox{\rm CS}}^{\mbox{\rm ind}}[\zeta]=
{b^\alpha\over 4\pi^2} A^\beta\tilde F_{\alpha\beta}
\left\{\zeta-1+\vartheta(-b^2-m^2)\sqrt{1+{m^2\over b^2}}\right\}\ . 
\label{5.27}
\end{equation}
The key point to be realized is that the one-loop radiatively 
induced CPT-odd Lagrange density is uniquely specified 
by a certain physical renormalization prescription.
Now, as a general inescapable feature of the presently investigated
modified QED, fermions of a given polarization and very high momenta
become unstable. The stability border in momentum space univocally
select the maximal residual symmetry subgroup of $O(3,1)^{++}$ which
remains to be the symmetry group at the quantum level in all 
the inertial frames in which the axial-vector $b_\mu$ has a given value. 
Consequently, the natural physical renormalization prescription at the 
perturbative level is such that the maximal residual symmetry is 
realized at the quantum level order-by-order in the loop expansion.
For instance, in the time-like case $b^2>0$,
the maximal symmetry group is reduced from $O(3,1)^{++}$ to $SO(3)$. 
In so doing, the freedom in the choice of the coefficient $\zeta$ 
reduces itself only to the freedom in the choice of the bare photon
CS vector $\eta_\mu$ in (\ref{2.1}).

As an alternative renormalization prescription, for instance,
one could require decoupling of fermions in the very large mass limit.  
This prescription looks physically quite reasonable, if the origin of the
radiatively induced CS term is not related to the heavy matter. 
This requirement can be  fulfilled by choosing
$\zeta=1$, up to the one-loop approximation. Then one finds that
the one-loop radiatively induced Lagrange density consistent with decoupling of
heavy fermions takes the form 
\begin{equation}
{\cal L}_{\mbox{\rm CS}}^{\mbox{\rm dec}}=
{b^\alpha\over 4\pi^2} A^\beta\tilde F_{\alpha\beta}
\vartheta(-b^2-m^2)
\sqrt{1+{m^2\over b^2}}\ , 
\label{5.28}
\end{equation}
a result which is in agreement with Refs.~\cite{3,18}. 
However, the instability of the
high-energy fermions of any mass in the presence of a constant 
axial-vector background - see Section 4 - does actually entail 
a different scenario, where
all fermions equally contribute and the induced CS vector accumulates
the effect of all charged fermions (see the next Section).
Conversely, the above decoupling may arise only as a result of a fine-tuning
between the net contribution of all fermions and the pure photon 
CS vector $\eta_\mu$.

A more detailed analysis of the origin of discrepancies between different
renormalization prescriptions and our scheme is presented in Appendix B.
Therein the role of maximal residual symmetry is elucidated to provide
the consistent definition of dimensional regularization. As well,
the different schemes are compared to, the leading order in the axial-vector
$b^\mu$, with respect to their prescription for angular integration. 
The latter one happens to be a key point to obtain different answers for the induced
CS coupling constant. 
\section{Conclusions: consistency between photons and fermions}
In this paper we have shown that the non-perturbative fermion
dynamics may be used to predict unambiguously the radiatively induced CS vector
\begin{equation}
\Delta\eta_\mu = {2\alpha \over\pi}\sum_{a=1}^N b^{a}_\mu,
\label{6.1}
\end{equation}
where the summation has to be performed over all the $N$ internal 
charged fermions degrees of freedom and the possibility to have different 
axial charges for different fermions is taken into account \cite{33}. 
>From the recent experimental bounds obtained in Refs.~\cite{27,28}
it is possible to estimate the magnitude of the spatial components 
of the induced CS vector to be
of the order $|\Delta{\vec \eta}|<10^{-19}\ \mbox{\rm eV}$, 
under the assumption that all the
vectors ${\vec b}^{a}$ are of the same order. 
If the vectors $b^{a}_\mu$ are related to some background torsion, 
then one expects them to be identical. 
If, however, they are generated 
by {\it e.g.} vacuum expectation values of gradients of axion fields,
then it is conceivable that $b^{a}_\mu$ might have different values.

On the one hand, the consistent quantization of fermions in a constant 
axial-vector field asks for the vectors $ b^{a}_\mu$ to be time-like  
- see Section 3 and Ref.~\cite{2} - which, nevertheless, does
not mean that the sum in eq.~(\ref{6.1}) is also time-like, because some of the 
fermions may have the opposite axial hypercharges.

On the other hand, the consistent quantization of photons can be achieved
when the full {\sl dressed} CS vector 
$\eta_\mu = \eta^{(0)}_\mu + \Delta\eta_\mu$ 
turns out to be  essentially purely space-like -
see Section 2 and Refs.~\cite{19,21}.
 
As we suppose that there
is some dynamical mechanism to generate -  
with the help of axion condensation - 
the purely bosonic part $\eta^{(0)}_\mu$ of the full CS four-vector,
the compatibility of the
consistent quantization of both fermions and bosons is believed to be
quite possible, contrary to the claim in Refs.~\cite{19,21}.

However, from the practical point of view, the cancellation between time components
of  $\eta^{(0)}_\mu$  and $\Delta\eta_\mu$ should be extremely
precise, in order to satisfy the experimental bounds \cite{1}-\cite{3} and
to fulfill the microcausality requirement of the photo-dynamics.
As well the severe experimental bounds on  $\vec b^{a}$ for electrons,
and protons \cite{27,28}, together with the estimation of birefringence
of radio-waves from remote galaxies and quasars, do not leave too much
room to eventually discover the CPT and Lorentz symmetry breaking 
in the quantum spinor and photon dynamics,
unless there is a striking cancellation among the addenda in eq.~(\ref{6.1}).
\acknowledgments{
It is a pleasure to thank Roman Jackiw, Alan Kosteleck\'y, Guy Bonneau,
Frans Klinkhamer and Manuel P\'erez-Victoria
for quite useful correspondences and remarks; we are also indebted to Lev Lipatov for 
an illuminating discussion. A. A. A. is partially supported by Grant RFBR
01-02-17152, Project INTAS 2000-587, Russian
Ministry of Education Grant E00-33-208 and by The Program {\sl Universities
of Russia: Fundamental Investigations} (Grant 992612); P. G. is supported
by Grant MURST-Cofin99.} 
\appendix
\section{Some identities in dimensional regularization} 
Here we list some useful identities concerning dimensional regularization.
The Levi-Civita symbol in the four dimensional Minkowski's space-time is normalized according to
\begin{equation}
\epsilon^{0123}=-\epsilon_{0123}\equiv 1\ ,
\label{A1}
\end{equation}
in such a way that the following identity holds true 
in the four dimensional Minkowski's space-time: namely,
\medskip
\begin{eqnarray}
&&\epsilon^{\mu\nu\alpha\beta}\epsilon_\mu^{\ \lambda\rho\sigma}=
\sum_{\{\nu\alpha\beta\}}(-)^P\ g^{\nu\lambda}g^{\alpha\rho}g^{\beta\sigma}=\nonumber\\
&&-
g^{\nu\lambda}g^{\alpha\rho}g^{\beta\sigma}-
g^{\alpha\lambda}g^{\beta\rho}g^{\nu\sigma}-
g^{\beta\lambda}g^{\nu\rho}g^{\alpha\sigma}\nonumber\\
&&+
g^{\nu\rho}g^{\alpha\lambda}g^{\beta\sigma}+
g^{\alpha\rho}g^{\beta\lambda}g^{\nu\sigma}+
g^{\beta\rho}g^{\nu\lambda}g^{\alpha\sigma}\ .
\label{A2}
\end{eqnarray}

\vspace{1.0cm}
\noindent
Concerning dimensional regularization, we collect here below the definitions
and key properties \cite{29} for the $2^\omega\times 2^\omega$ $\gamma$-matrices in a $2\omega$-dimensional
space-time with Minkowski signature
\begin{eqnarray}
\gamma^\mu&=& \bar\gamma^\mu, \quad \mu=0,1,2,3;\nonumber\\
 &=&\hat\gamma^\mu,\quad \mu=4,\ldots ,2\omega-4.
\label{A3}
\end{eqnarray}
\medskip
\begin{equation}
\left\{\bar\gamma^\mu,\bar\gamma^\nu\right\}=2\bar g^{\mu\nu}
{\bf 1}_{2^\omega}\ ;\quad
\left\{\hat\gamma^\mu,\hat\gamma^\nu\right\}=2\hat g^{\mu\nu}
{\bf 1}_{2^\omega}\ ;\quad
\left\{\bar\gamma^\mu,\hat\gamma^\nu\right\}=0\ .
\label{A4}
\end{equation}
\medskip
\begin{equation}
\left\|\bar g\right\|={\tt diag}(+,-,-,-)\ ;\qquad 
\left\|\hat g\right\|=(-){\bf
1}_{2\omega-4}\ ;
\label{A5} 
\end{equation}
\medskip
\begin{equation}
\gamma_5\equiv i\bar\gamma^0\bar\gamma^1\bar\gamma^2\bar\gamma^3\ ;\quad
\gamma_5^2={\bf 1}_{2^\omega}\ ;\quad
\left\{\bar\gamma^\mu,\gamma_5\right\}=0=
\left[\hat\gamma^\mu,\gamma_5\right]\ . 
\label{A6}
\end{equation}
\bigskip
\noindent
Taking all the above listed equations into account, it is not difficult to
check the following trace formulae: {\it i.e.}, 
\begin{eqnarray}
&&{\tt tr}\left(\gamma_5\bar\gamma^\mu\bar\gamma^\lambda\bar\gamma^\rho
\bar\gamma^\nu\right)=-i2^\omega\epsilon^{\mu\lambda\rho\nu}\ ; \label{A7}\\
&&{\tt tr}\left(\gamma_5\bar\gamma^\mu\bar\gamma^\lambda\bar\gamma^\rho
\bar\gamma^\nu\bar\gamma^\sigma\bar\gamma^\tau\right)=\nonumber\\
&&i2^\omega\left(
\epsilon^{\nu\sigma\tau\mu}\bar g^{\lambda\rho}+
\epsilon^{\nu\sigma\tau\rho}\bar g^{\lambda\mu}+
\epsilon^{\mu\lambda\rho\sigma}\bar g^{\nu\tau}\right)\nonumber\\
&&-i2^\omega\left(
\epsilon^{\nu\sigma\tau\lambda}\bar g^{\mu\rho}+
\epsilon^{\mu\lambda\rho\nu}\bar g^{\sigma\tau}+
\epsilon^{\mu\lambda\rho\tau}\bar g^{\nu\sigma}\right)\ . \label{A8}
\end{eqnarray}
\section{Comparison of different renormalization prescriptions} 
Let us now clarify the origin of the discrepancy between different
perturbative renormalization prescriptions and our scheme.  
First, we comment about the $b_\mu$-linear contribution of the 
one-loop radiatively induced Lagrange density\footnote{The 
one-loop radiatively induced Lagrange density to the leading-order in
$b_\mu$ has been recently evaluated and discussed by many
authors using different regularization procedures and getting different 
results. For a recent review and references see \cite{20}.}
within the framework of dimensional regularization.
As a matter of fact, it appears from eq.~(\ref{5.25}) that the relevant quantity
$\Pi^{\mu\nu}_{\mbox{\rm odd}}(0;b,m)$  to lowest order in $b_\mu$ is non-vanishing 
if dimensional regularization is employed. At  first
sight, this result seems to contradict the calculation of Ref.~\cite{18} leading instead
to a vanishing result for the corresponding quantity. The resolution of 
this puzzling feature is quite instructive.

The discrepancy does actually originate, at the perturbative level, from 
a subtlety in the very definition of the dimensionally regularized one-loop 
self-energy (\ref{5.1}). In the present context, the key point is the inversion of
the classical kinetic differential operator of eq.~(\ref{4.5}). We recall that the
fermion propagator (\ref{4.6}) - which coincides with the one of Ref.~\cite{14} -  has been obtained by 
inverting the classical kinetic differential operator in terms of the conventional 
four-dimensional Clifford's algebra. Then, once inserted in the regularized expression (\ref{4.3}), 
the $\gamma$-matrices are understood in $2\omega$-dimensions according to 
the 't Hooft-Veltman-Breitenlohner-Maison consistent algebraic rules - see \cite{29} and Appendix A. 
This dimensional regularization scheme will be referred
to as $\overline{DR}$.

Alternatively, one can first extend the classical spinor kinetic operator onto $2\omega$ dimensions. 
Then the inverse operator will give the following Feynman's propagator:
\begin{equation}
(-i)\tilde S(p)={p^2+ b^2-m^2+2\left(b\cdot
p+m \not\!b\right)\gamma_5 -  [\not\!b, \not\!{\hat p}]\gamma_5\over
\left(p^2+b^2-m^2+i\varepsilon\right)^2-4\left[(b\cdot
p)^2- b^2\left(m^2 -\hat p^2\right)\right]}\left( \not\!p
+m+\not\!b\gamma_5\right)\ , 
\label{B1}
\end{equation}
which differs from the propagator of eq.~(\ref{4.6}) by the last term in the numerator and 
the last term in the denominator respectively. It is worthwhile to draw the attention 
to the fact that this possible alternative does indeed produce some additional and 
unwanted Lorentz symmetry breaking, as the mass-shell and pole structure of the propagator (\ref{4.6}) is spoiled. 
For instance, in the case of time-like $b_\mu$ the symmetry group is not an invariance group of the vector $b_\mu$ itself,
{\it i.e.} $O(2\omega -1)$, but instead $O(3)\times O(2\omega - 4)$. 

Therefore, this alternative implementation $\widehat{DR}$ of the dimensional regularization 
for the one-loop self-energy based on the propagator (\ref{B1}) actually leads to some
extra unphysical source for the Lorentz symmetry breaking. On the contrary, the use 
of the propagator (\ref{4.6}) just provides the maximal residual 
symmetry $O(2\omega -1)$ of the mass-shell compatible with a
Lorentz symmetry breaking due to the presence of a background axial-vector.
It explains the precise agreement between 
the physical cutoff and $\overline{DR}$ calculations,
because the $\overline{DR}$ scheme manifestly
keeps the maximal residual space-time
symmetry in the enlarged $D$-dimensional space-time
and, consequently, it does provide the very same 
analytical structure on the complex energy plane.

As already emphasized, the alternative implementation $\widehat{DR}$ of the dimensional 
regularization based on the propagator (\ref{B1}) can only 
affect $b_\mu$-linear contribution of the one-loop self-energy.
When taking the propagator (\ref{B1}) at the linear order in  $b_\mu$
\begin{equation}
\tilde S(p)=S(p) - {i[\not\!b, \not\!{\hat p}]\gamma_5\over
\left(p^2-m^2+i\varepsilon\right)^2}\left(\not\!p +m\right)\ , 
\label{B2}
\end{equation}
it is not difficult to check that the contribution of the second term 
in the RHS completely compensates the contribution of the first one
into the $b_\mu$-linear part of the radiatively induced CPT-odd vertex.
In so doing, the Lagrange density of eq.~(\ref{5.28}) is eventually recovered and 
the vanishing result of Refs.~\cite{3,18} is endorsed at the leading-order in $b_\mu$. 
However, we stress once again that this conclusion is achieved at the price of some 
additional Lorentz symmetry breaking which
drastically changes the mass-shell structure coming from
the classical fermion kinetic differential operator in eq.~(\ref{4.5}). 
Owing to this main reason we give our favor to the choice (\ref{4.5}) 
for the fermion propagator. 
It leads to the one-loop perturbative result (\ref{5.25}), 
which is consistent with the existence of the physical cutoff.

Finally,  we would like again to emphasize the technical point 
where the discrepancy arises between
different regularization approaches. It precisely corresponds to the 
integration of the term involving $4p_\rho (b\cdot p)$ in eq.~(\ref{4.9}). 
In Refs.~\cite{13,14} the assumption of the exact Lorentz
symmetry of  4-momentum space was adopted, {\it i.e.}, the averaging relation
$\left<4 p_\rho (b\cdot p)\right>_{D=4} = b_\rho p^2$ was employed, where the
average obviously indicates angular integration. 
However, this symmetry is certainly broken and only the rotational 
3-momentum symmetry survives for the time-like $b^\mu$. 
As a consequence, the average with respect to the spatial  3-momentum  yields
$\left< 4 p_\rho (b\cdot p)\right>_{D=3} = 4 b_0 p^2_0\delta_{\rho 0}$, 
the result  which 
has been used in eq.~(\ref{4.9}).
The same result in the $\overline{DR}$ scheme has been achieved
thanks to  the identity $\left<4 p_\rho (b\cdot p)\right>_{D=2\omega} = 
2 b_\rho p^2/\omega$, together with subsequent derivation of the simple pole 
in $(2-\omega)$.

\end{document}